\documentstyle[twocolumn,aps,psfig]{revtex}

\begin{document}
\draft
\flushbottom
\twocolumn[
\hsize\textwidth\columnwidth\hsize\csname @twocolumnfalse\endcsname

\title{ Nonlinear optics of surface plasmon toy black holes.}
\author{Igor I. Smolyaninov}
\address{ Department of Electrical and Computer Engineering \\
University of Maryland, College Park,\\
MD 20742}
\date{\today}
\maketitle
\tightenlines
\widetext
\advance\leftskip by 57pt
\advance\rightskip by 57pt

\begin{abstract}
Nonlinear optical behavior of recently introduced surface plasmon toy black hole models has been considered. Physical realization of these models involves droplets of liquid on a metal surface, which supports propagation of surface plasmons. Such droplets are shown to exhibit strongly enhanced nonlinear optical behavior in the frequency range near the surface plasmon resonance of a metal-liquid interface. This enhancement may be responsible for the missing orders of magnitude of field enhancement in the surface enhanced Raman scattering effect. In addition, experimental observation of the recently predicted optical second harmonic generation near the toy event horizon is reported. Finally, the possibility of toy Hawking radiation observation is discussed, and the expression for the effective Hawking temperature of a toy surface plasmon black hole is derived. The effective Hawking temperature appears to be in the range of a few hundreds Kelvin for the realistic droplet shapes, which may lead to a surprising conclusion that the shapes of the liquid droplets on metal substrates are to some degree determined by the Hawking radiation.  
\end{abstract}

]
\narrowtext

\tightenlines

The realization that solid-state toy models may help in an understanding of electromagnetic phenomena in curved space-time has led to considerable recent effort in developing toy models of electromagnetic \cite{1} and sonic \cite{2} black holes. In the case of media electrodynamics this is possible because of an analogy between the propagation of light in matter and in curved space-times: it is well known that the Maxwell equations in a general curved space-time background $g_{ik}(x,t)$ are equivalent to the phenomenological Maxwell equations in the presence of a matter background with nontrivial electric and magnetic permeability tensors $\epsilon _{ij}(x,t)$ and $\mu _{ij}(x,t)$ \cite{3}. In this analogy, the event horizon corresponds to a surface of singular electric and magnetic permeabilities, so that the speed of light goes to zero, and light is "frozen" near such a surface. In the absence of established quantum gravitation theory the toy models are helpful in understanding electromagnetic phenomena in curved space-times, such as Hawking radiation \cite{4} and the Unruh effect \cite{5}. Unfortunately, until recently all the suggested electromagnetic black hole toy-models were very difficult to realize and study experimentally, so virtually no experimental work was done in this field. 

Very recently a simple surface plasmon toy black hole model has been suggested and observed in the experiment \cite{6}. In addition, a surface plasmon toy model of a rotating black hole (Kerr metric \cite{7}) has been introduced \cite{8}. These models are based on the surface plasmon resonance at the edge of a liquid droplet deposited on the metal surface, which supports propagation of surface plasmons. Surface plasmons are collective excitations of the conductive electrons and the electromagnetic field \cite{9}. They exist in "curved three-dimensional space-times" defined by the shape of the metal-dielectric interface. Since in many experimental geometries surface plasmons are weakly coupled to the outside world (to free-space photons, phonons, single-electron excitations, etc.) it is reasonable to treat the physics of surface plasmons separately from the rest of the surface and bulk excitations, so that a field-theory of surface plasmons in a curved space-time background may be considered. For example, a nanohole in a thin metal membrane may be treated as a "wormhole" connecting two "flat" surface plasmon worlds located on the opposite interfaces of the membrane \cite{10}. On the other hand, near the plasmon resonance (which is defined by the condition that $\epsilon _m(\omega )=-\epsilon _d$ at the metal-dielectric interface, where $\epsilon _m(\omega )$ and $\epsilon _d$ are the dielectric constants of metal and dielectric, respectively \cite{9}) the surface plasmon velocity vanishes, so that the surface plasmon "stops" on the metal surface, and the surface charge and the normal component of the electric field diverge. For a given frequency of light, the spatial boundary of the plasmon resonance ("the event horizon" of our toy model) may be defined at will using the geometry $\epsilon _d(x,y)$ of the absorbed layer of dielectric on the metal surface. Thus, the plasmon resonance becomes a natural candidate to emulate the event horizon of a black hole. As a result, toy two-dimensional surface plasmon black holes can be easily produced and studied \cite{6,8}. 

In what follows I am going to explore the nonlinear optical behavior of the surface plasmon "black holes" based on the droplets of liquid on the gold or silver surfaces. I will show that such droplets exhibit strongly enhanced nonlinear optical behavior in the frequency range near the surface plasmon resonance of a metal-liquid interface due to considerable optical field enhancement near the toy event horizon (the edge of a droplet). This enhancement may be responsible for the missing orders of magnitude of field enhancement in the surface enhanced Raman scattering effect \cite{11}. In addition, experimental observation of the recently predicted \cite{12} optical second harmonic generation near the toy event horizon will be reported. Finally, the possibility of toy Hawking radiation observation will be discussed, and the expression for the effective Hawking temperature of a toy surface plasmon black hole will be derived.

As a first step, let us consider the basic features of the surface plasmon toy black hole analogy. Let us consider in detail the dispersion law of a surface plasmon (SP), which propagates along the metal-dielectric interface. The SP field decays exponentially both inside the metal and the dielectric. Inside the dielectric the decay exponent is roughly equal to the SP wave vector. As a first step let us assume that both metal and dielectric completely fill the respective $z<0$ and $z>0$ half-spaces. In such a case the dispersion law can be written as \cite{9} 

\begin{equation}  
k^2=\frac{\omega ^2}{c^2}\frac{\epsilon _d\epsilon _m(\omega )}{\epsilon _d+\epsilon _m(\omega)} ,
\end{equation}

where we will assume that $\epsilon _m=1-\omega _p^2/\omega ^2$ according to the Drude model, and $\omega _p$ is the plasma frequency of the metal. This dispersion law is shown in Fig.1(b) for the cases of metal-vacuum and metal-dielectric interfaces. It starts as a "light line" in the respective dielectric at low frequencies and approaches asymptotically $\omega =\omega _p/(1+\epsilon _d)^{1/2}$ at very large wave vectors. The latter frequency corresponds to the so-called surface plasmon resonance. Under the surface plasmon resonance conditions both phase and group velocity of the SPs is zero, and the surface charge and the normal component of the electric field diverge. Since at every wavevector the SP dispersion law is located to the right of the "light line", the SPs of the plane metal-dielectric interface are decoupled from the free-space photons due to the momentum conservation law.   

If a droplet of dielectric (Fig.1(a)) is placed on the metal surface, the SP dispersion law will be a function of the local thickness of the droplet. Deep inside the droplet far from its edges the SP dispersion law will look similar to the case of a metal-dielectric interface, whereas near the edges (where the dielectric is thinner) it will approach the SP dispersion law for the metal-vacuum interface. As a result, for every frequency between $\omega _p/(1+\epsilon _d)^{1/2}$ and $\omega _p/2^{1/2}$ there will be a closed linear boundary inside the droplet for which the surface plasmon resonance conditions are satisfied. Let us show that such a droplet of dielectric on the metal interface behaves as a "surface plasmon black hole" in the frequency range between $\omega _p/(1+\epsilon _d)^{1/2}$ and $\omega _p/2^{1/2}$, and that the described boundary of the surface plasmon resonance behaves as an "event horizon" of such a black hole. 

Let us consider a SP within this frequency range, which is trapped near its respective "event horizon", and which is trying to leave a large droplet of dielectric (see Fig.2). The fact that the droplet is large means that ray optics may be used. Since the component of the SP momentum parallel to the droplet boundary has to be conserved, such a SP will be totally internally reflected by the surface plasmon resonance boundary back inside the droplet at any non-zero angle of incidence. This is a simple consequence of the fact that near the "event horizon" the effective refractive index of the droplet for surface plasmons is infinite (according to eq.(1), both phase and group velocity of surface plasmons is zero at surface plasmon resonance). On the other hand, even if the angle of incidence is zero, it will take the SP infinite time to leave the resonance boundary. Thus, the droplet behaves as a black hole for surface plasmons, and the line near the droplet boundary where the surface plasmon resonance conditions are satisfied plays the role of the event horizon for surface plasmons. 

The toy black holes described above are extremely easy to make and observe \cite{6}. In our experiments a small droplet of glycerin was placed on the gold film surface and further smeared over the surface using lens paper, so that a large number of glycerin microdroplets were formed on the surface (Fig.3(a)). These microdroplets were illuminated with white light through the glass prism (Fig.1(a)) in the so-called Kretschman geometry \cite{9}. The Kretschman geometry allows for efficient SP excitation on the gold-vacuum interface due to phase matching between the SPs and photons in the glass prism. As a result, SPs were launched into the gold film area around the droplet. Photograph taken under a microscope of one of such microdroplets is shown in Fig.3(b). The white rim of light near the edge of the droplet is clearly seen. It corresponds to the effective SP event horizon described above. Near this toy event horizon SPs are stopped or reflected back inside the droplet. In addition, a small portion of the SP field may be scattered out of the two-dimensional surface plasmon world into normal three-dimensional photons. These photons produced the image in Fig.3(b). We also conducted near-field optical measurements of the local surface plasmon field distribution around the droplet boundary Fig.3(c,d,e) using a sharp tapered optical fiber as a microscope tip. These measurements were performed similar to the measurements of surface plasmon scattering by individual surface defects described in \cite{13}. The droplet was illuminated with 488 nm laser light in the Kretschman geometry. The tip of the microscope was able to penetrate inside the glycerin droplet, and measure the local plasmon field distribution both inside and outside the droplet. Inside the droplet (in the right half of the images) the shear-force image (c) corresponds to the increase in viscous friction rather than the droplet topography. However, this image accurately represents the location of the droplet boundary, shown by the arrow in Fig.3(e). The sharp and narrow local maximum of the surface plasmon field just inside the droplet near its boundary is clearly visible in the near-field image Fig.3(d) and its cross-section Fig.3(e).         

Unfortunately, the described toy SP black hole model does not work outside the frequency range between $\omega _p/(1+\epsilon _d)^{1/2}$ and $\omega _p/2^{1/2}$. On the other hand, this is a common feature of every electromagnetic toy black hole model suggested so far. All such toy models necessarily work only within a limited frequency range. In addition, the losses in metal and dielectric described by the so far neglected imaginary parts of their dielectric functions will put a stop to the singularities of the field somewhere very near the toy event horizon (however, such "good" metals as gold and silver have very small imaginary parts of their dielectric constants, and hence, very pronounced plasmon resonances \cite{9}). Notwithstanding these limitations, the ease of making and observing such toy SP black holes makes them a very promising research object. If we forget about the language of "black holes" and "event horizons" for a moment, the SP optics phenomenon represented in Fig.3(b) remains a potentially very interesting effect in surface plasmon optics. Namely, this photo shows the existence of a two-dimensional SP analog of whispering gallery modes, which are well-known in the optics of light in droplets and other spherical dielectric particles. Whispering gallery modes in liquid microdroplets are known to substantially enhance nonlinear optical phenomena due to cavity quantum electrodynamic effects \cite{14}. One may expect even higher enhancement of nonlinear optical mixing in liquid droplets on the metal surfaces due to enhancement of surface electromagnetic field inherent to surface plasmon excitation, and in addition, due to accumulation of SP energy near the surface plasmon event horizons at the droplet boundaries. This strong enhancement of nonlinear optical effects in liquid droplets may be very useful in chemical and biological sensing applications. It may also be responsible for the missing orders of magnitude of field enhancement in the surface enhanced Raman scattering (SERS) effect \cite{11}, since various plasmon excitations are believed to play a major role in SERS. 

The current explanation of SERS is based on the combination of electromagnetic and "chemical" enhancements \cite{11}. The current calculations of the electromagnetic field enhancement take into account "electrostatic" field enhancement at the apexes of various surface protrusions (the "lightning rod effect"), and the local field enhancement due to excitation of various localized surface plasmon modes in the crevices of the rough metal film. In addition, various weak and strong surface plasmon localization effects are considered in combination with the consideration of a rough metal surface as a fractal object \cite{15}. The "chemical" enhancement was proposed as an explanation for the quite a few missing orders of magnitude in theoretically calculated optical field enhancement, which follows from the magnitude of the experimentally measured SERS signals \cite{11}. The "chemical" enhancement may happen if the molecular energy levels are affected by the proximity to the rough metal surface, and are drawn into resonance with the excitation field. 

SERS observations are usually conducted under not so well controlled conditions when the surface topography is not well-known and "dirty" (the target molecules are present on the surface in random locations). On the other hand, it is well known that even monolayer surface coverages considerably shift plasmon resonance \cite{9} of the metal-vacuum interface. It would not be a big stretch to suggest that some areas of the "dirty" metal surface may contain compact areas covered with the multiple layers of the target or solvent molecules (even if there are no droplets on the surface). Such compact areas would affect surface plasmon propagation in a way, which is very similar to the effect of the droplets considered above: effective surface plasmon event horizons may appear near the boundaries of such areas. Experimental data shown in Fig.3 and the theoretical arguments above strongly indicate that the local field enhancement near these boundaries may be considerable. Even the data shown in Fig.3(e) obtained with the limited optical resolution of the order of 100 nm \cite{13} indicate at least 10-fold enhancement of the square of the local field intensity near the droplet boundary. In the physical picture of surface plasmon whispering gallery modes the local SERS signal may be expected to grow by a factor of $Q^2$, where $Q$ is the cavity quality factor \cite{14}. For the surface plasmons in the visible range $Q$ may be estimated roughly as $Q\sim L/2\pi R$, where $L$ is the surface plasmon free propagation length and $R$ is the droplet radius. Taking into account the typical theoretical value of $L\sim 40\mu m$ \cite{16} in the visible range, $Q^2\sim 50$ may be obtained for a $R=1\mu m$ droplet. Thus, such surface plasmon "whispering galleries"/"black holes" may provide considerable SERS enhancements on top of the local electromagnetic enhancement due to other effects associated with the surface roughness.

The effects of surface roughness on the properties of plasmons near the effective event horizon have not been considered so far. However, detailed analyses of these effects conducted recently for the case of real black holes \cite{12} indicates that these effects will be especially strong in the nonlinear optics domain. For example, strong optical second harmonic generation has been predicted near the event horizon due to the effects of weak localization \cite{12}. 

Wave propagation and localization phenomena in random media have been the topic of extensive studies during the last years \cite{17}. One of the most striking examples of such phenomena is the strong and narrow peak of diffuse second harmonic light emission observed in the direction normal to a randomly rough metal surface (see Fig.4(a)). This peak is observed under the coherent illumination at any angle. This effect was initially predicted theoretically \cite{18} and later observed in the experiment \cite{19}. The enhanced second harmonic peak normal to the mean surface arises from the fact that a state of momentum {\bf k} introduced into a weakly localized system will encounter a significant amount of backscattering into states of momentum centered about {\bf -k}. When these surface {\bf k} and {\bf -k} modes of frequency $\omega $ interact through an optical nonlinearity to generate $2\omega $ radiative modes, the $2\omega $ light has nonzero wave vector components only perpendicular to the mean surface. The angular width of the normal peak can be as small as a few degrees, and its amplitude far exceeds the diffuse omnidirectional second harmonic background. 

Similar weak localization effects may be expected for the surface plasmon modes, which are trapped near the effective event horizon of our toy black hole when the metal surface exhibits moderate roughness. Similar to the case of planar rough surface \cite{18,19}, the momentum component parallel to the edge of the droplet must be conserved in nonlinear optical processes. This means that while the surface plasmon {\bf k} and {\bf -k} modes of frequency $\omega $, which interact through an optical nonlinearity, generate $2\omega $ modes, the $2\omega $ plasmons could have nonzero wave vector component only perpendicular to the mean edge of the droplet (Fig.4(b)). Thus, weak localization effects in the scattering of surface optical modes near the horizon should produce a pronounced peak in the angular distribution of second harmonics of plasmons in the direction normal to the horizon. In addition, because of such propagation direction (perpendicular to the edge of the horizon), these second harmonic plasmons have the best chances to escape the vicinity of the toy black hole. As a result, the relative intensity of the second harmonic radiation due to the weak localization effect near the event horizon will be much higher with respect to the diffuse SHG than in the case of planar rough surface. The diffuse SH plasmons would remain trapped near the horizon similar to the case of a real black hole \cite{12}.

Our experiments strongly indicate enhancement of SHG near the toy surface plasmon black holes. The droplets shown in Fig.5(a) were illuminated by the weakly focused beam (illuminated spot diameter on the order of 50 $\mu $m) from a Ti:sapphire laser system consisting of an oscillator and a regenerative amplifier operating at 810 nm (repetition rate up to 250 kHz, 100-fs pulse duration, and up to 10 $\mu $J pulse energy), which was directed onto the sample surface in the Kretschman geometry (Fig.1(a)). Excitation power at the sample surface was kept below the ablation threshold of the gold film. The local SHG from the two droplets illuminated by the laser can be clearly seen in Fig.5(b), which was obtained using a far-field microscope and a CCD camera. It should be noted, however, that while SH frequency plasmons experience the "event horizon" near the edges of the droplets, the refractive index of glycerine is not sufficiently large for the fundamental plasmons to see it. Thus, 810 nm plasmons may only be trapped into the whispering gallery modes near the edges of the droplets. On the other hand, the basic weak localization mechanism remains the same in the studied experimental situation.

Finally, let us examine the potentially most interesting nonlinear optical phenomenon, which may be exhibited by our toy black holes. The Hawking radiation \cite{4} has been one of the most fascinating quantum physical phenomena discovered recently. However, nobody has observed it in the experiment so far, and its observation remains a great experimental challenge. In order to evaluate the chances to see it in the experiments performed with surface plasmon toy black holes, we must first derive the expression for the effective Hawking temperature. A derivation of an effective Hawking temperature of a toy dielectric black hole has been performed recently by Reznik \cite{1}. Below we will closely follow his procedure. 

In order to emulate the effective space-time geometry considered in \cite{1}, let us consider a thin metal membrane with two "linear" droplets positioned symmetrically on both sides of the membrane (Fig.6(a)). The droplets thicknesses at $x=0$ correspond to the surface plasmon resonance at the illumination frequency. The droplets taper off adiabatically in positive and negative $x$-directions on both sides of the membrane. In the symmetric membrane geometry the surface plasmon spectrum consists of two branches $\omega _-$ and $\omega _+$, which exhibit positive and negative dispersion, respectively, near the surface plasmon resonance \cite{20}. Fig.6(b) shows the dispersion curves of both branches for the cases of metal-vacuum interface far from the droplets, and for the locations near $x=0$. The droplets represent an effective black hole for $\omega _-$ modes and an effective white hole for $\omega _+$ modes, similar to \cite{1}. If the effective slowly varying local surface plasmon phase velocity $c^\star (x) =\omega /k(x)$ is introduced, the wave equation for surface plasmons may be written as 

\begin{equation}
(\frac{\partial ^2}{c^{\star 2}\partial t^2}-\frac{\partial ^2}{\partial x^2}-\frac{\partial ^2}{\partial y^2})A=0,
\end{equation}

which corresponds to an effective metric

\begin{equation}
ds^2=c^{\star 2}dt^2-dx^2-dy^2
\end{equation}

The event horizon corresponds to $c^\star =0$ at $x=0$. The two regions $x>0$ and $x<0$ correspond to the two sides of a black hole, which are connected by an Einstein-Rosen bridge at $x=0$. The behavior of $c^\star (x)$ near $x=0$ may be defined at will by choosing the corresponding geometry of the droplet edge (if necessary, the droplet may be replaced by a similar shaped layer of solid dielectric). In order to adhere to the metric considered in \cite{1} let us assume that $c^\star =\alpha xc$ in the vicinity of $x=0$. However, we should remember that this linear behavior will be cut off somewhere near the effective horizon due to such effects as Landau damping, losses in the metal and the dielectric, etc. \cite{6,8}. The resulting effective metric now looks like

\begin{equation}
ds^2=\alpha ^2x^2c^2dt^2-dx^2-dy^2
\end{equation}

Thus, this particular choice of the shape of the droplet edge gives rise to an effective Rindler geometry. Similar metric would describe the space-time geometry near the event horizon of a black hole with mass $M_{BH}=c^2/8\gamma \alpha $, where $\gamma $ is the gravitation constant \cite{1}. 

Following the standard derivation procedure \cite{1,2,4,5}, we must now introduce the analogue "Minkowski" coordinates as $U=-xe^{-\alpha t}$ and $V=xe^{\alpha t}$, which make the effective metric above conformally flat, and require that in the flat "Minkowski" space the plasmon field will be in its vacuum state $\mid 0_M>$. However, the "Minkowski" vacuum looks like a bath of thermal radiation with the temperature

\begin{equation}
T_H=\frac{M_{PL}^2c^2}{8\pi k_BM_{BH}}=\frac{\hbar c\alpha }{\pi k_B},
\end{equation}

where $M_{PL}=(\hbar c/\gamma )^{1/2}$ is the Planck mass and $k_B$ is the Boltzmann constant. For more details on the derivation of this result one may address \cite{1}, where it is also demonstrated that cutting off $c^\star $ near $x=0$ and inclusion of the dispersion does not eliminate the Hawking radiation. 

We should point out that the final result for the effective Hawking temperature does not depend on the gravitation constant $\gamma $. This is an encouraging fact since we are dealing with the toy black holes. In order to get a numerical estimate on the effective Hawking temperature we must assume some realistic value of $\alpha $. Our approximation of a slow adiabatically changing $c^\star $ inside the droplet means that $c^\star $ does not change considerably on the scale of the local wavelength $\lambda ^\star $ of surface plasmons. Thus, a good top estimate for $\alpha $ should look like $\alpha <1/\lambda \sim 1/\lambda _0$, where $\lambda $ is the plasmon wavelength far from the droplet, and $\lambda _0$ is the wavelength of light in vacuum. As a result, the effective Hawking temperature of our toy black hole is of the order of 

\begin{equation}
T_H\leq \frac{1}{2\pi ^2}\frac{\hbar \omega }{k_B}\sim 1000K,
\end{equation} 

where $\omega $ is the characteristic surface plasmon frequency.
This value is quite close to the room temperature, which means that the droplet may probably conform itself to be exactly in thermal equilibrium with its ambient. Thus, we come to a surprising conclusion that the shape of liquid droplets on metal surfaces may to some degree be determined by the Hawking radiation. 

However, we should accept this result with a degree of caution because of the recent claim that the electromagnetic black hole analogues exhibit only the classical features of the black holes, while such quantum mechanical properties as Hawking radiation may not be reproduced \cite{21}. The basic reason for not having thermal radiation emitted is the energy conservation: there should be a source of thermal energy emitted to infinity. On the other hand, we must remember that the surface plasmons are non-radiative modes, which are bound to the interface. In addition, they do not propagate far alone the interface, being absorbed by the metal within a few tens of micrometers from the source. As a result, the surface plasmon Hawking radiation associated with the toy black hole would be observable only in the near field of the black hole, and energy conservation would not be violated. This situation would be rather similar to the recently discovered near-field thermal surface phonon-polariton emission in SiC \cite{22}, when the monochromatic emission with photon energies not represented in the far field zone may be observed in the near field of the SiC surface. In fact, this interesting effect is caused by the fact that $c^\star \rightarrow 0$ under the conditions of surface phonon-polariton resonance, similar to the toy Hawking radiation effect considered above. In addition, one may create non-equilibrium situations when energy is pumped into a toy black hole. Under such non-equilibrium conditions Hawking radiation may also be observable in the experiment. 

In conclusion, nonlinear optical behavior of recently introduced surface plasmon toy black hole models has been considered. Physical realization of these models involves droplets of liquid on a metal surface, which supports propagation of surface plasmons. Such droplets are shown to exhibit strongly enhanced nonlinear optical behavior in the frequency range near the surface plasmon resonance of a metal-liquid interface. This enhancement may be responsible for the missing orders of magnitude of field enhancement in the surface enhanced Raman scattering effect. In addition, experimental observation of the recently predicted optical second harmonic generation near the toy event horizon has been reported. Finally, the possibility of toy Hawking radiation observation has been discussed, and the expression for the effective Hawking temperature of a toy surface plasmon black hole has been derived.

This work has been supported in part by the NSF grants ECS-0210438 and ECS-0304046.

Figure captions.

Fig.1 (a) Experimental geometry of surface plasmon toy black hole observation. (b) Surface plasmon dispersion law for the cases of metal-vacuum and metal-dielectric interfaces

Fig.2 A surface plasmon trapped inside a droplet near the effective "event horizon": The projection of surface plasmon momentum parallel to the droplet edge must be conserved. Due to effectively infinite refractive index near the droplet edge surface plasmons experience total internal reflection at any angle of incidence. 

Fig.3 Far-field (a,b) and near-field (c,d) images of a toy surface plasmon black hole: (a) Droplet of glycerin on a gold film surface (illuminated from the top). The droplet diameter is approximately 15 micrometers. (b) The same droplet illuminated with white light in the Kretschman geometry, which provides efficient coupling of light to surface plasmons on the gold-vacuum interface (Fig.1(a)). The white rim around the droplet boundary corresponds to the effective surface plasmon "event horizon". (c) and (d) show $10\times 10 \mu m^2$ topographical and near-field optical images of a similar droplet boundary (droplet is located in the right half of the images) illuminated with 488 nm laser light. Cross-sections of both images are shown in (e). Position of a droplet boundary is indicated by the arrow.  

Fig.4 (a) Strong enhancement of second harmonic emission from a randomly rough metal surface in the direction normal to the surface. A typical angular distribution of the diffuse second harmonic light is shown in the inset. (b) Second harmonic generation near the toy black hole horizon. Surface roughness is represented by a rough droplet edge. Second harmonic plasmons emitted perpendicular to the horizon have the best chances to escape the toy black hole.

Fig.5 (a) Microscopic image of the glycerin droplets under normal illumination. (b) SHG is seen from the droplets illuminated by 810 nm Ti:sapphire laser light in the Kretschman geometry shown in Fig.1(a). 

Fig.6 (a) A thin metal membrane with two "linear" droplets positioned symmetrically on both sides of the membrane. An effective event horizon is located at $x=0$. Such geometry emulates the effective space-time geometry considered in \cite{1}. (b) Dispersion laws of the $\omega _-$ and $\omega _+$ modes exhibit positive and negative dispersion, respectively, near the surface plasmon resonance. These branches are shown for the cases of metal-vacuum interface far from the droplets, and for the locations near $x=0$.  


\begin{references}

\bibitem{1} B. Reznik, Phys.Rev.D 62, 044044 (2000).

\bibitem{2} W.G. Unruh, Phys.Rev.D 51, 2827 (1995).

\bibitem{3} W. Schleich and M.O. Scully, in New Trends in Atomic Physics, edited by G. Grynberg and R. Stora (North Holland, Amsterdam, 1984) p.997.

\bibitem{4} S. Hawking, Nature 248, 30 (1974).

\bibitem{5} W.G. Unruh, Phys.Rev.D 14, 870 (1976).

\bibitem{6} I.I. Smolyaninov and C.C. Davis, gr-qc/0306089, submitted for publication in Phys.Rev.Lett.

\bibitem{7} L.D. Landau and E.M. Lifshitz, Field Theory (Pergamon, New York, 1984).

\bibitem{8} I.I. Smolyaninov, cond-mat/0309356, submitted for publication in the New Journal of Physics.

\bibitem{9} H. Raether, Surface Plasmons, Springer Tracts in Modern Physics Vol.111 (Springer, Berlin, 1988).

\bibitem{10} I.I. Smolyaninov, Phys.Rev.B 67, 165406 (2003). 

\bibitem{11} A.G. Malshukov, Phys.Reports 194, 343 (1990).

\bibitem{12} I.I. Smolyaninov, Phys.Letters A 300, 375-380, (2002).

\bibitem{13} I.I. Smolyaninov, D.L. Mazzoni and C.C. Davis, Phys.Rev.Letters, 77, 3877 (1996). 

\bibitem{14} A.J. Campillo, J.D. Eversole, and H.B. Lin, Phys.Rev.Lett. 67, 437 (1991).

\bibitem{15} V.M. Shalaev, Nonlinear Optics of Random Media (Springer, Berlin, 2000).

\bibitem{16} P. Dawson, F. de Fornel, and J.-P. Goudonnet, Phys.Rev.Lett. 72, 2927 (1994).

\bibitem{17} A. Ishimaru, Wave Propagation and Scattering in Random Media (IEEE, New York, 1997).

\bibitem{18} A.R. McGurn, T.A. Leskova, and V.M. Agranovich, Phys.Rev.B 44, 11441 (1991).

\bibitem{19} O.A. Aktsipetrov, V.N. Golovkina, O.I. Kapusta, T.A. Leskova, and N.N. Novikova, Phys.Lett.A 170, 231 (1992).

\bibitem{20} K.L. Kliewer and R. Fuchs, Phys.Rev. 144, 495 (1966).

\bibitem{21} W.G. Unruh and R. Schutzhold, Phys.Rev. D 68, 024008 (2003).

\bibitem{22} A.V. Shchegrov, K. Joulain, R. Carminati, and J.-J. Greffet, Phys.Rev.Letters 85, 1548 (2000).

\end{references}
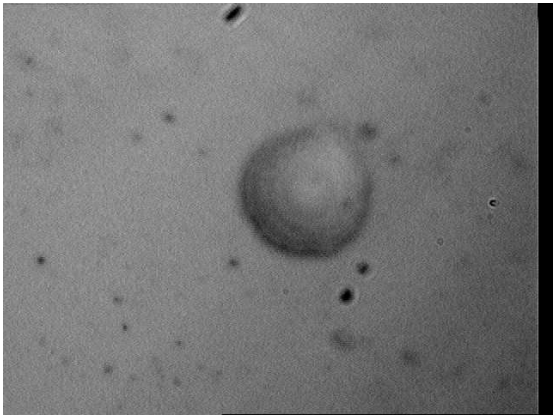
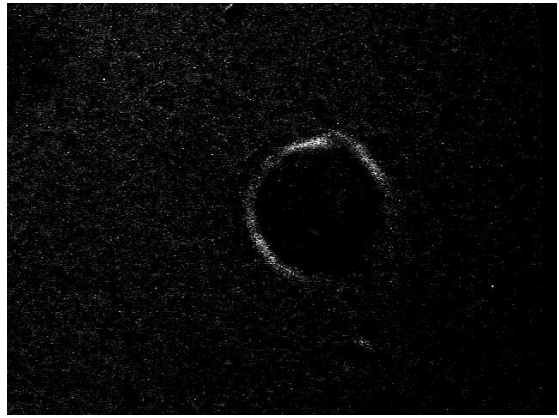
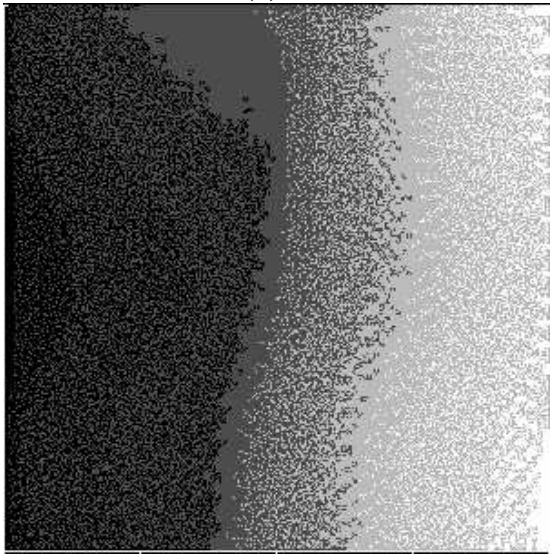
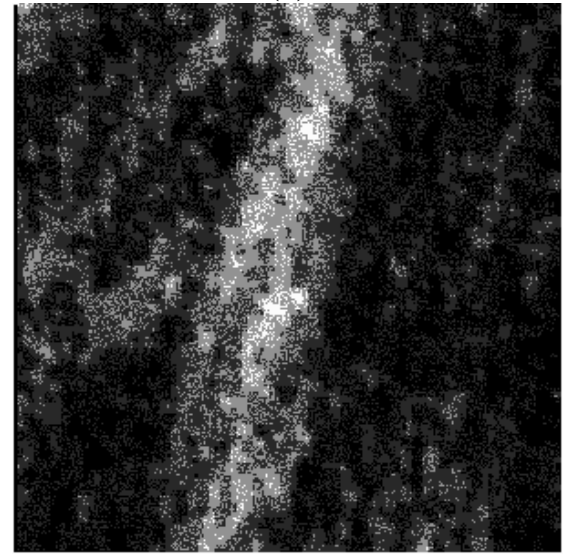
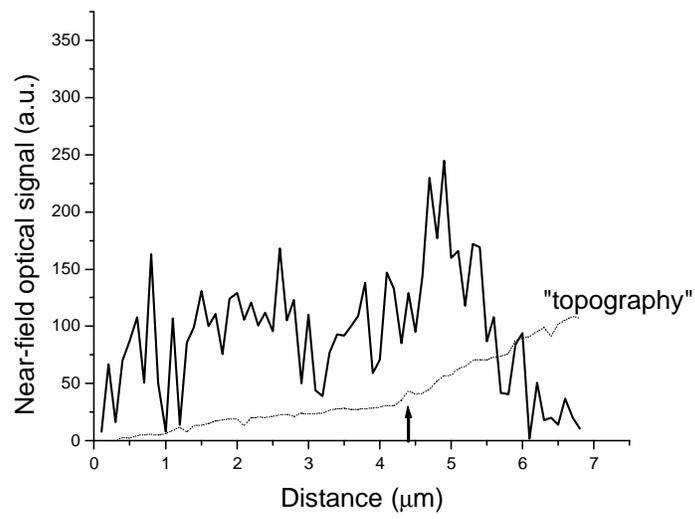
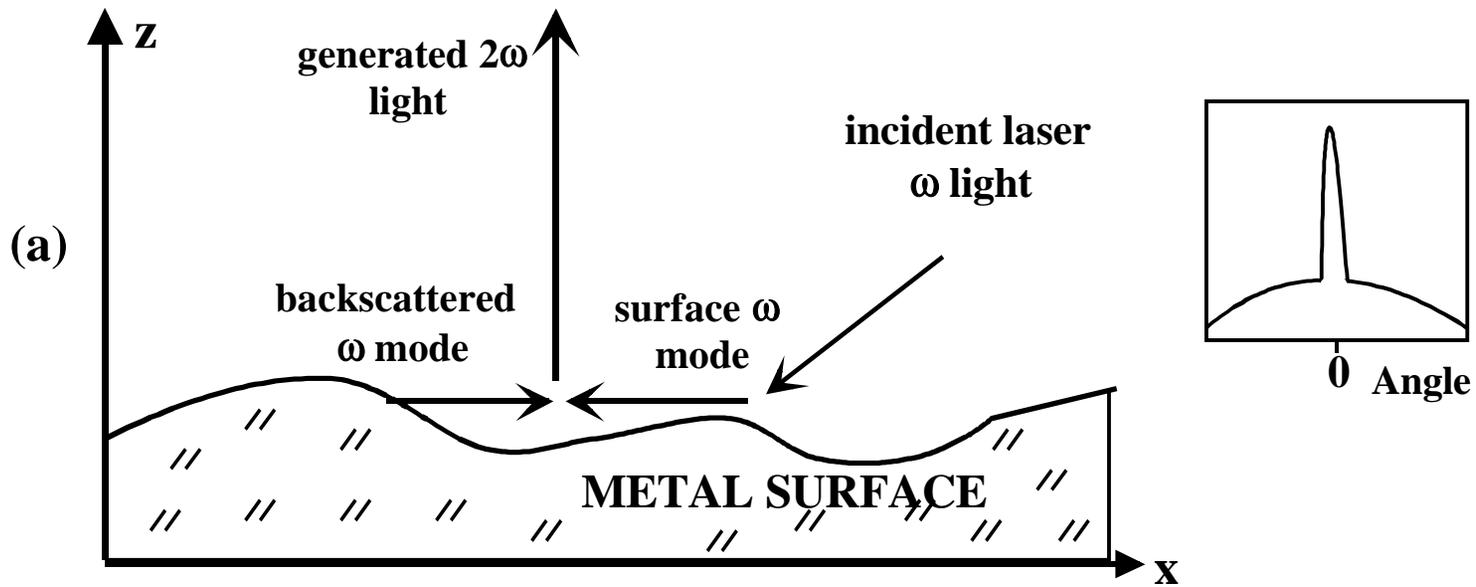
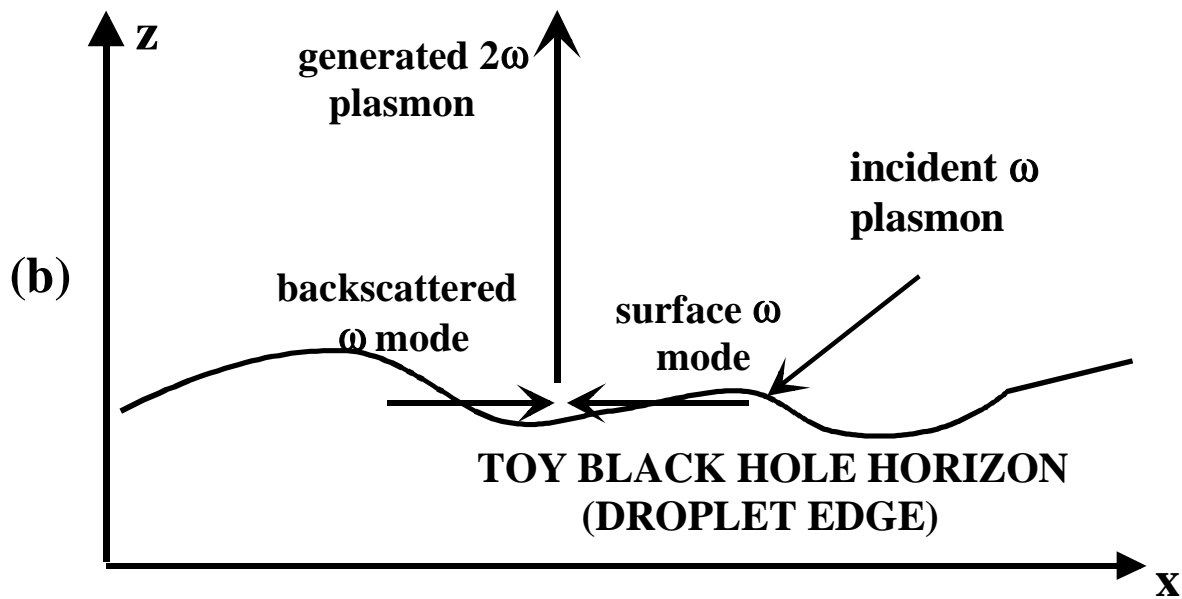
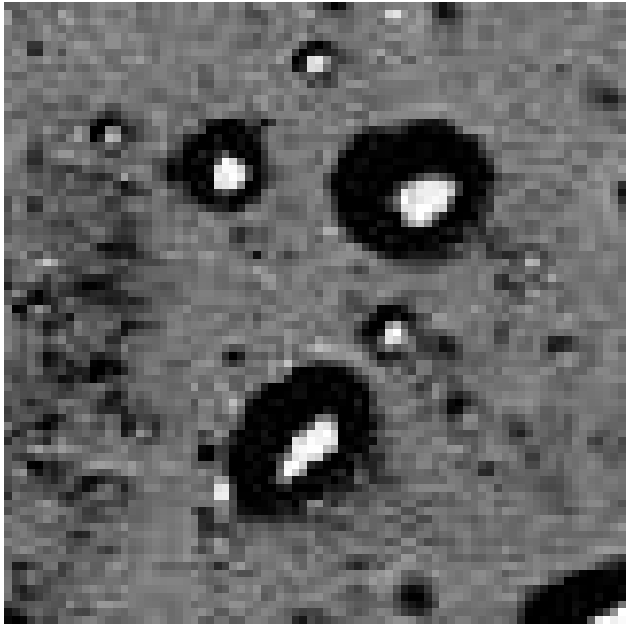
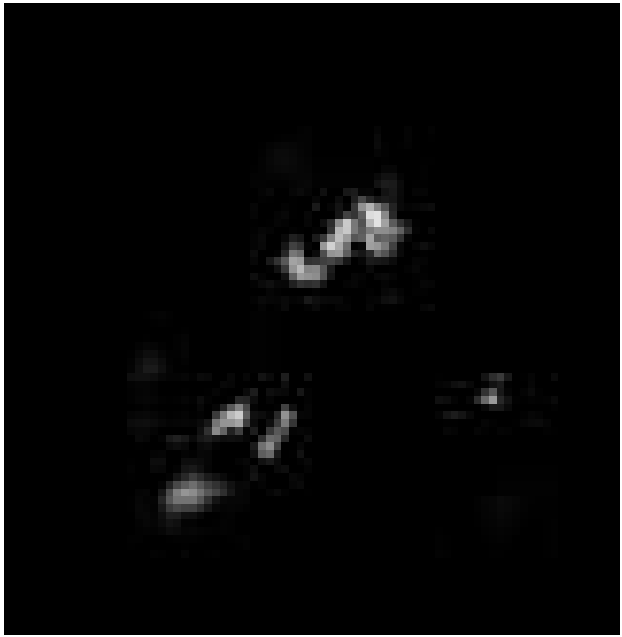
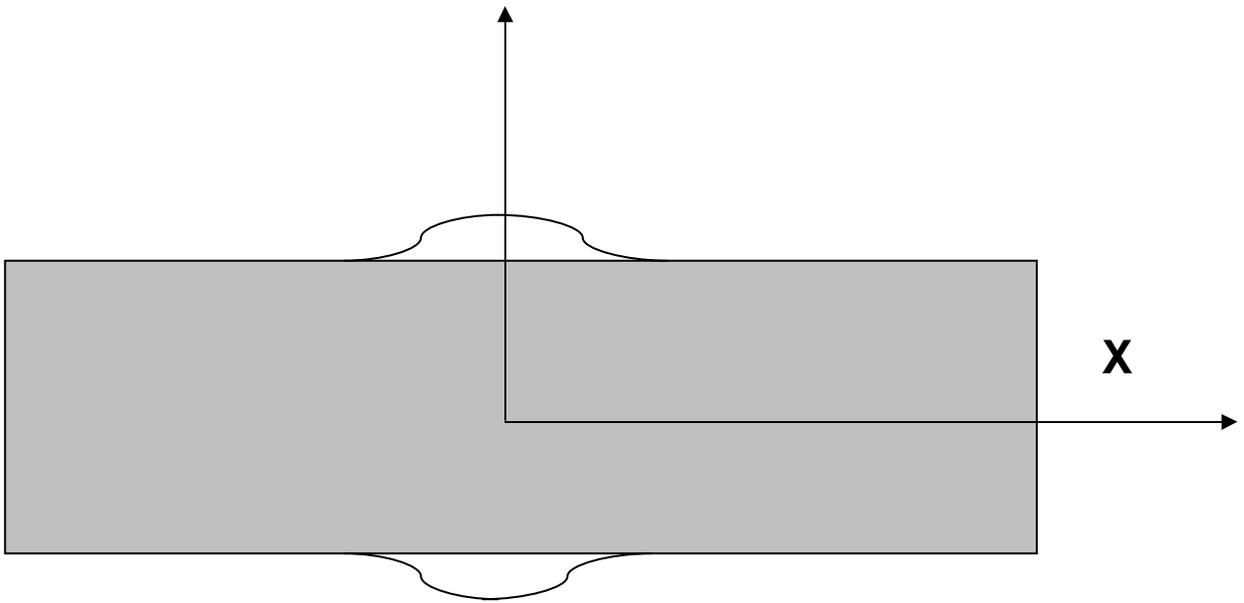
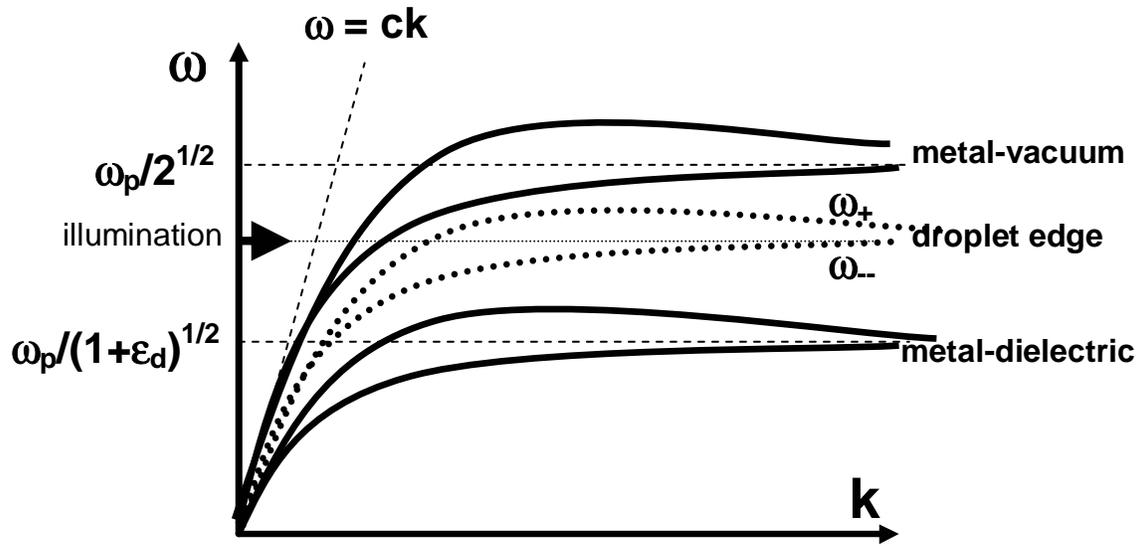
\end{document}